\pdfoutput=1

\documentclass[11pt]{article}
\usepackage{tcolorbox}
\DeclareUnicodeCharacter{FF1A}{:}

\usepackage{acl}
\usepackage{bbm}
\usepackage{times}
\usepackage{latexsym}

\usepackage{amssymb}   

\usepackage[T1]{fontenc}

\usepackage[utf8]{inputenc}

\usepackage{amssymb}    
\usepackage{colortbl}   
\usepackage{xcolor}     
\usepackage{array}      
\usepackage{booktabs}    

\usepackage{hyperref}
\usepackage{url}
\usepackage{booktabs}       
\usepackage{times}  
\usepackage{helvet}  
\usepackage{courier}  
\usepackage{graphicx}
\usepackage{algorithm}
\usepackage{algpseudocode}
\usepackage{amsmath}
\usepackage{multirow}
\usepackage{tabularx}
\usepackage{colortbl}
\usepackage{array}
\usepackage{marvosym}
\usepackage{xcolor}
\usepackage{natbib}  
\usepackage{caption} 
\usepackage{wrapfig}
\usepackage{adjustbox}
\usepackage{lipsum}
\usepackage{soul}
\usepackage{pifont}      
\usepackage{xcolor}      
\definecolor{lightgray}{gray}{0.7}

\usepackage{subfigure}
\usepackage{stfloats}    
\usepackage{placeins} 
\usepackage{listings}

\newcommand{\eat}[1]{}

\newcommand{\OurDATA}{\textsc{CodeMEnv}}

\title{\OurDATA{}: Benchmarking Large Language Models on Code Migration}

\author{
Keyuan Cheng\textsuperscript{*,1,3,4},
Xudong Shen\textsuperscript{*,1,4},
Yihao Yang\textsuperscript{*,1,4},
Tengyue Wang\textsuperscript{1,4},
Yang Cao\textsuperscript{1,4},\\
\textbf{Muhammad Asif Ali\textsuperscript{2},
Hanbin Wang\textsuperscript{3},
Lijie Hu\textsuperscript{†,1,2},
Di Wang\textsuperscript{†,1,2}}\\
$^1$Provable Responsible AI and Data Analytics (PRADA) Lab\\
$^2$King Abdullah University of Science and Technology \\
$^3$Peking University \quad
$^4$South China University of Technology
}

\begin{document}

\maketitle

\begin{abstract}
Large language models (LLMs) have shown remarkable capabilities across various software engineering tasks; however, their effectiveness in code migration—adapting code to run in different environments—remains insufficiently studied. In this work, we introduce \OurDATA{}: \underline{\textbf{Code}} \underline{\textbf{M}}igration Across \underline{\textbf{Env}}ironment, a new benchmark specifically designed to assess LLMs' abilities in code migration scenarios. \OurDATA{} consists of 922 examples spanning 19 Python and Java packages, and covers three core tasks: 
(1) identifying functions incompatible with specific versions, 
(2) detecting changes in function definitions, and 
(3) adapting code to target environments. 
Experimental evaluation with seven LLMs on~\OurDATA{} yield an average pass@1 rate of 26.50\%, with \textsc{GPT-4o} achieving the highest score at 43.84\%. Key findings include: (i) LLMs tend to be more proficient with newer function versions, which aids in migrating legacy code, and (ii) LLMs sometimes exhibit logical inconsistencies by identifying function changes irrelevant to the intended migration environment. The datasets are available at \url{https://github.com/xdshen-ai/Benchmark-of-Code-Migration}.
\end{abstract}
\def\thefootnote{*}\footnotetext{Equal Contribution.}
\def\thefootnote{†}\footnotetext{Corresponding Author.}

\section{Introduction}
\label{sec:intro}
Large Language Models (LLMs) have demonstrated remarkable abilities in software engineering tasks, including code generation~\cite{Jiang2024ASO,Du2024EvaluatingLL} and code translation~\cite{Yuan2024TRANSAGENTAL,Eniser2024TowardsTR}. General-purpose LLMs such as GPT-4~\cite{Achiam2023gptGPT4TR}, Claude-3~\cite{TheC3}, and DeepSeek~\cite{Shao2024DeepSeekV2AS} consistently achieve state-of-the-art results across diverse programming challenges, often outperforming traditional approaches on established benchmarks. Beyond these general models, a range of specialized CodeLLMs have been introduced to further advance performance on code-related tasks. Notable examples include CodeT5+~\cite{Wang2023CodeT5OC}, CodeLlama~\cite{Rozire2023CodeLO}, and StarCoder2~\cite{Lozhkov2024StarCoder2A}. Thanks to their tailored architectures and training data, these models exhibit a deep understanding of code structure and syntax, frequently surpassing general LLMs on programming-specific benchmarks.

\begin{figure}
    \centering
    \includegraphics[width=1\linewidth]{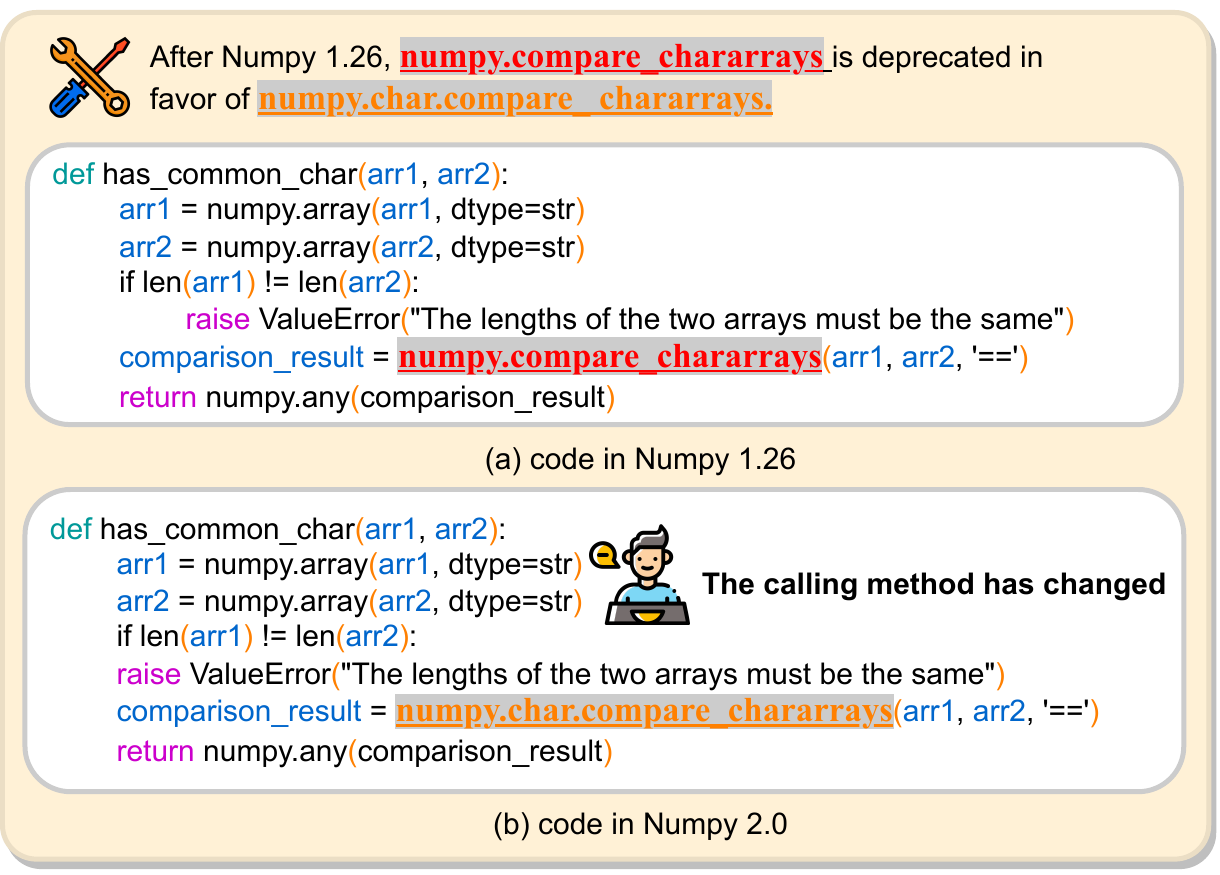}
    \caption{The function $\texttt{compare\_chararrays}$ underwent changes after Numpy 1.26, creating compatibility issues between NumPy 1.26 and 2.0.}
    \vspace{-3.7ex}
    \label{fig:intro}
\end{figure}

Despite the impressive progress of LLMs in various code-related tasks, their ability to perform code migration—adapting code to run correctly in a different environment—remains largely unexplored. Code migration is a critical challenge in practical software development. For example, when users attempt to run code obtained from sources like GitHub, they often encounter compatibility issues that require significant manual effort to resolve. If LLMs could automate the migration of code to fit a user's existing environment, it would greatly streamline the process of environment setup and reduce the burden of manual configuration.

The root cause of these compatibility issues lies in the ongoing evolution and versioning of software libraries. As libraries are updated and maintained, APIs and function calls may change, leading to incompatibilities across different environments. For instance, as illustrated in Figure~\ref{fig:intro}, the function call \texttt{compare\_chararrays} was moved from the \texttt{numpy} package to the \texttt{numpy.char} submodule. This change results in functionally equivalent code being implemented differently in NumPy 1.26 versus NumPy 2.0, highlighting the challenges of code migration across library versions.

\eat{
Research in this area is still in its early stages, with only a few studies exploring the challenges and potential solutions.
Some benchmarks, like CodeUpdate~\cite{Liu2024CodeUpdateArenaBK}, focus on how to inject knowledge of new API functions that the model has never seen before. 
However, the functions used in these studies are synthetically generated by GPT rather than sourced from real-world libraries. This limitation makes it challenging to assess the feasibility of code migration in real-world scenarios.
}

Research on code migration is still in its infancy, with the majority of prior work concentrating on cross-language migration (i.e., code translation) rather than migration across different library versions or environments. Only a few recent efforts, such as that by Google researchers~\cite{Ziftci2025MigratingCA}, have explored automated LLM-based approaches for identifying code changes required for cross-version migration. However, there remains a significant gap: the absence of comprehensive benchmarks to systematically assess the code migration capabilities of LLMs.

\begin{figure}
    \centering
    \includegraphics[width=\linewidth]{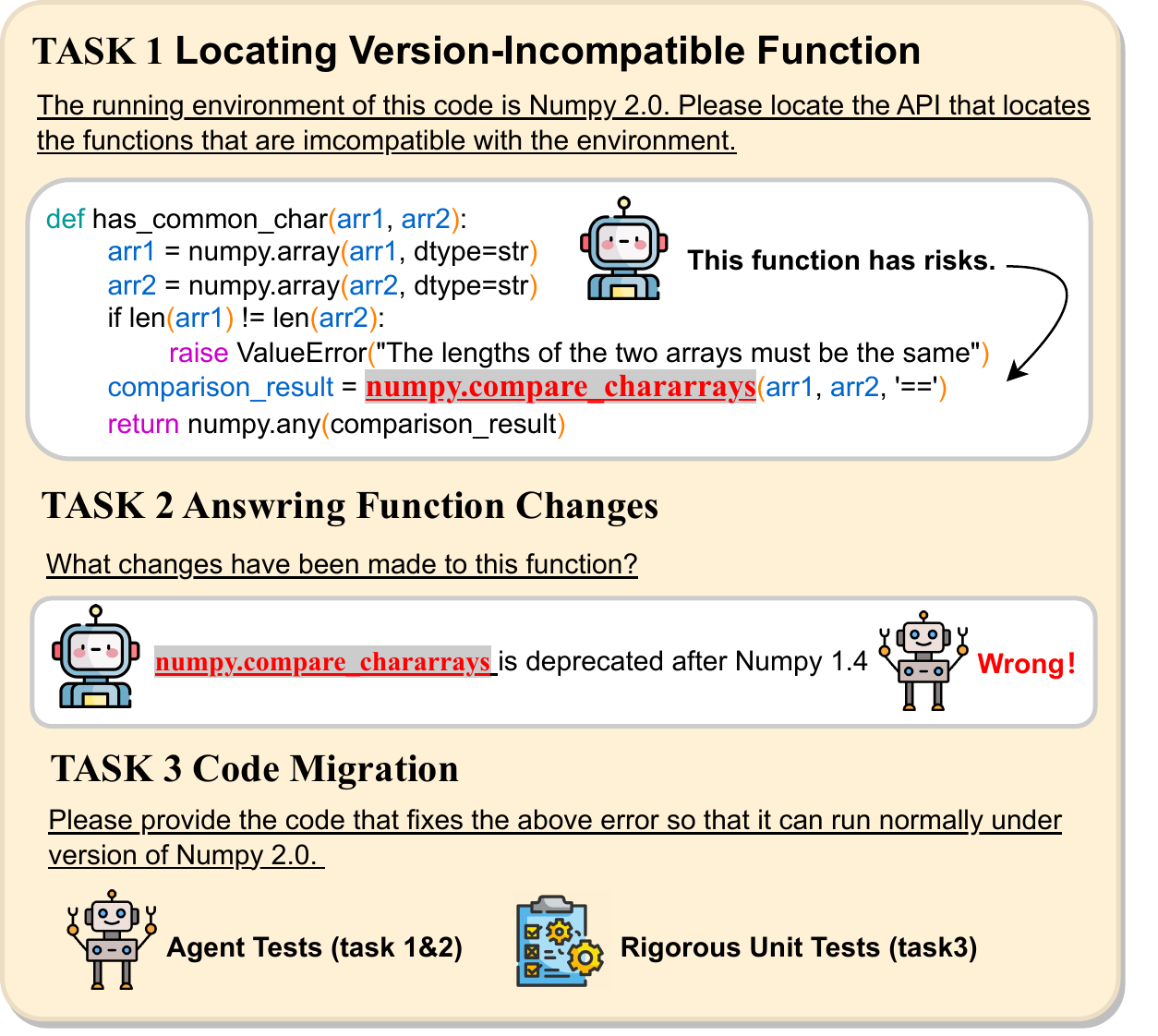}
    \caption{A data example of \OurDATA{}, which includes three tasks to evaluate LLMs on environment-related programming skills.}
    \label{fig:task}
    \vspace{-2.7ex}
\end{figure}

To address this gap, we introduce a new benchmark, \OurDATA{} (\underline{\textbf{Code}} \underline{\textbf{M}}igration Across \underline{\textbf{Env}}ironment). \OurDATA{} is built from 922 real-world function changes, manually curated from official sources, spanning 19 Python and Java packages. As shown in Figure~\ref{fig:task}, the benchmark is organized into three tasks that collectively evaluate LLMs' abilities to perform code migration:

\noindent \textbf{Task 1: Locate version-incompatible functions.} Given a code snippet and a specified target environment, the model is tasked with identifying functions or code segments that may not be compatible with the running environment.\\
\noindent \textbf{Task 2: Answering function changes.} The model must explain the specific modifications these functions have undergone between versions.\\
\noindent \textbf{Task 3: Code migration.} The model is then required to revise or migrate the provided code to ensure it runs correctly in the target environment, addressing any version-related incompatibilities.

To evaluate~\OurDATA{}, we conduct experiments with nine LLMs. Results show that, for the migration task, the average pass@1 rate across seven models is 26.50\%, with \textsc{GPT-4o} achieving the highest score at 43.84\%. Our analysis uncovers several notable insights:

\noindent \textbf{(i) Familiarity with function versions.} LLMs tend to be more familiar with newer function versions, which enables them to migrate legacy code to modern environments more effectively, but makes adapting newer code to older environments more challenging.

\noindent \textbf{(ii) Logical inconsistencies.} LLMs sometimes display logical inconsistencies when identifying relevant function changes for migration. For example, when migrating code from version 1.16 to 2.0, models may mistakenly reference changes from version 1.0, which are not pertinent to the migration at hand.

\vspace{-1.7ex}
\section{Related Work}
\vspace{-1.7ex}
\label{sec:RW}

\subsection{LLMs for Code}

Large Language Models (LLMs) have achieved remarkable progress in code-related tasks such as generation, translation, and completion, owing to their large parameter counts and training on vast code corpora. Proprietary models like GPT-4~\cite{Achiam2023gptGPT4TR}, Claude-3~\cite{TheC3}, and Gemini~\cite{Reid2024Gemini1U} consistently deliver strong performance across a broad spectrum of programming challenges. In parallel, open-source models such as Qwen2.5~\cite{qwen2.5} have demonstrated competitive or even superior results compared to larger models, leveraging synthetic data to enhance their capabilities. Other open-source models, including Llama-3.1~\cite{Dubey2024TheL3}, Phi-3~\cite{Abdin2024Phi3TR}, and DeepSeek~\cite{Shao2024DeepSeekV2AS}, also achieve impressive results, with DeepSeek in particular outperforming many proprietary alternatives.

The field has also seen rapid advances in specialized CodeLLMs. For example, CodeT5~\cite{Wang2021CodeT5IU} employs an encoder-decoder architecture and excels at code completion and retrieval, while CodeT5+~\cite{Wang2023CodeT5OC} introduces flexible encoder-only, decoder-only, and encoder-decoder modes, achieving strong results on benchmarks like HumanEval. CodeLlama~\cite{Rozire2023CodeLO}, developed by Meta, extends Llama 2 with code-specific training and supports multiple languages, with its 34B variant showing robust performance in code generation and completion. StarCoder2~\cite{Lozhkov2024StarCoder2A}, from BigCode, is designed for multi-language code synthesis and retrieval, leveraging large-scale permissively licensed datasets. Qwen-Coder~\cite{Hui2024Qwen25CoderTR} is notable for its 32B variant, which surpasses GPT-4o on several benchmarks, benefiting from training on 5.5T tokens of diverse data and strong human preference alignment. These developments underscore the rapid evolution of domain-specific LLMs and the narrowing gap between open-source and proprietary solutions.

In this work, we assess the LLMs' ability to migrate code across different environments.

\vspace{-1.2ex}
\subsection{Code Migration}
\vspace{-1.2ex}
Recent progress in AI-driven code migration has demonstrated encouraging results across diverse programming scenarios. Amazon Q Developer~\cite{AmazonQ} exemplifies a generative AI tool tailored to assist developers in upgrading Java applications from versions 8/11 to 17, addressing the broader challenges of repository-level code migration. Joe~\cite{Joe2024migration} provides a comprehensive analysis of the current landscape and persistent obstacles in large-scale migration efforts. The dynamics of human-AI collaboration in this context are explored by Omidvar Tehrani et al.~\cite{OmidvarTehrani2024EvaluatingHP}, who assess how developers and Amazon Q Developer interact during migration tasks. Google researchers~\cite{Ziftci2025MigratingCA} have introduced an automated approach that integrates change location discovery with LLM-based guidance to streamline migration processes. In the domain of legacy system modernization, Kontogiannis~\cite{Kontogiannis2010CodeMT} proposes a semi-automated method for migrating PL/IX to C++, emphasizing iterative optimization to address performance issues. More recently, Almeida et al.~\cite{Almeida2024AutomaticLM} demonstrated that ChatGPT can effectively support API migration, with One-Shot prompting proving particularly effective for migrating Python/SQLAlchemy applications while preserving functionality and adopting modern features.

Additional discussion of related work is provided in Appendix~\ref{relatedApp}.




\begin{figure*}
    \centering
    \includegraphics[width=1\linewidth]{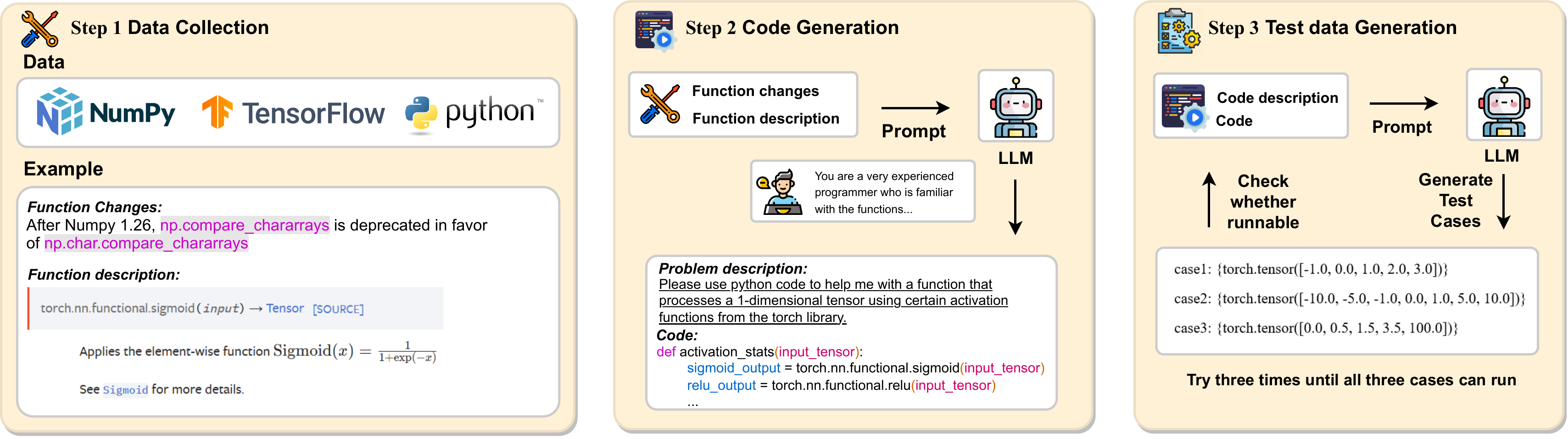}
    \caption{The construction process of \OurDATA{}. \textbf{Step 1:} We collect function change information and function descriptions from the official website; \textbf{Step 2:} Based on the collected functions, generate code that can run in the original version and its problem description; \textbf{Step 3:} Generate 3 test cases for each data and repeat three times until all cases can run correctly.}
    \label{fig:process}
    \vspace{-1.7ex}
\end{figure*}
\vspace{-1.7ex}
\section{\OurDATA{}}
\vspace{-1.7ex}
\label{sec:data_curate}
We argue that despite significant challenges that environment-related issues present to programmers, there is currently no systematic benchmark for evaluating model capabilities in code migration across different environments. To fill this gap, we introduce \OurDATA{} (\underline{\textbf{Code}} \underline{\textbf{M}}igration Across \underline{\textbf{Env}}ironments), a benchmark designed to assess models' understanding of function usage differences across library versions and their ability to perform cross-version code migration. The remainder of this section details the construction and characteristics of \OurDATA{}.

\vspace{-1.7ex}
\subsection{Task Definition}
\OurDATA{} include the following three tasks:

\noindent {\bf Task-1: Identify Version-Incompatible Functions.} Given a code snippet and a specified target environment version, the model is asked to pinpoint functions that are incompatible with that environment. This task is divided into two levels of difficulty: \textbf{easy}, where only one incompatible function is present, and \textbf{hard}, where multiple incompatible functions must be identified.

\noindent {\bf Task-2: Answering Function Changes.} For each function identified in Task-1 the model must explain the nature of the change. This includes specifying the type of change (e.g. deprecation), the version in which the change occurred, and any replacement function if applicable.

\noindent {\bf Task-3: Code Migration.} The model is required to revise the provided code so that it is compatible with the target environment. Code migration scenarios are further categorized as follows:

\noindent {\bf (a) \textsc{New2Old}:} The target environment version is older than the original, so the model must adapt newer code to run in a legacy environment.

\noindent {\bf (b) \textsc{Old2New}:} The target environment version is newer than the original, requiring the model to upgrade legacy code for compatibility with the updated environment.

\subsection{Dataset Statistics}

\OurDATA{} encompasses two widely used programming languages in the deep learning community: Python and Java. The Python portion of the dataset spans 11 packages and contains a total of 587 samples, which are divided into two difficulty levels: \textbf{easy} and \textbf{hard}. The \textbf{easy} subset comprises 396 samples, each featuring a single line of code that is incompatible with the target environment. The \textbf{hard} subset includes 191 samples, each containing $k$ incompatible lines of code, where $k \in \{2, 3\}$. Table~\ref{tab:dataset-single-column} summarizes the distribution of incompatible lines across the datasets.

For Java, the dataset covers 8 packages with 335 samples. Only the \textbf{easy} difficulty level is included for Java, as the incompatible functions in these packages tend to be loosely connected, making it difficult to construct \textbf{hard} instances with multiple interdependent incompatibilities.

Additional details and comprehensive statistics for~\OurDATA{} are provided in Appendix~\ref{statistics}.

\subsection{Function Changes}
\label{changes}
We categorize function changes in \OurDATA{} into three main types:
\begin{itemize}
\itemsep-0.5em 
    \item \textbf{Addition} ($\texttt{None}\rightarrow f_{\text{new}}$): A new function $f_{\text{new}}$ is introduced in a later version, meaning it is unavailable in earlier environments.
    \item \textbf{Deprecation} ($f_{\text{old}}\rightarrow \texttt{None}$): The function $f_{\text{old}}$ is removed or no longer supported after a certain version, so it cannot be used in newer environments.
    \item \textbf{Replacement} ($f \rightarrow f^\prime$): The function $f$ is replaced or modified to $f^\prime$, which may involve changes to the function name, parameters, or usage patterns.
\end{itemize}
Table~\ref{tab:functionchanges} in Appendix~\ref{statistics} summarizes the distribution of these change types in the Python portion of \OurDATA{}: 98 replacements, 35 deprecations, and 79 additions.

\begin{table}[ht]
\centering
\resizebox{\linewidth}{!}{ 
\begin{tabular}{lllll}
\toprule[1.0pt]
\textbf{Datasets}  & \textbf{1-incom.} & \textbf{2-incom.} & \textbf{3-incom.} & \textbf{Total}\\
\midrule
\multirow{1}{*}{\textbf{Python (easy)}}  &396  &-  &- &396 \\
\multirow{1}{*}{\textbf{Python (hard)}}  &-  &103  &88 &191 \\  
\multirow{1}{*}{\textbf{Java}}   &335 &-  &- &335 \\
\bottomrule[1.0pt]
\end{tabular}}
\caption{Statistics of the number of incompatible lines of \OurDATA{}.}
\label{tab:dataset-single-column}
\vspace{-3.7ex}
\end{table}

\subsection{Evaluation}
\label{metrics}
In this section, we outline the evaluation methodology for the three benchmark tasks, utilizing two primary approaches:

\noindent {\bf Agent-based Evaluation.}
To evaluate whether LLMs can accurately identify version-incompatible functions and correctly describe the changes those functions have undergone, we adopt an agent-based evaluation strategy. The agent is provided with ground-truth answers, including the set of incompatible functions and their corresponding changes. It then compares the LLMs' predictions to these references according to the following criteria:

For Task-1, the evaluation requires the model to identify all functions that are incompatible with the target environment. The prediction is considered correct only if the set of identified functions exactly matches the ground truth; missing or incorrectly including any function results in failure. The accuracy for Task-1 is defined as:
\begin{equation}
\mathrm{Acc}_{\text{Task-1}} = \mathbbm{1}\left[ \mathcal{I} = \mathcal{T} \right] = 
\begin{cases}
1, & \text{if } \mathcal{I} = \mathcal{T} \\
0, & \text{otherwise}
\end{cases},
\end{equation}
where $\mathcal{I}$ is the set of ground-truth incompatible functions, and $\mathcal{T}$ is the set predicted by the LLM.

For Task-2, we assess three aspects:
(i) whether the LLM correctly identifies the type of change (see Section~\ref{changes});
(ii) whether the predicted version number of the change is accurate (allowing a margin of 0.5 between predicted and actual version numbers);
(iii) for replacement-type changes, whether the LLM correctly specifies the replacement function (this is not required for addition or deprecation cases).
The accuracy for Task-2 is given by:
\begin{equation}
\begin{aligned}
\mathrm{Acc}_{\text{Task-2}} = \mathbbm{1}[\, &\underbrace{\hat{t} = t}_{\text{type}} \land \underbrace{|\hat{v} - v| \leq 0.5}_{\text{version}} \\
&\land\, \underbrace{(t \neq \text{`Replace'} \lor \hat{f} = f)}_{\text{function}}\,],
\end{aligned}
\end{equation}
where $\hat{t}$ and $t$ are the predicted and ground-truth change types, $\hat{v}$ and $v$ are the predicted and actual version numbers, and $\hat{f}$ and $f$ are the predicted and ground-truth replacement functions.

Further details on the agent-based evaluation are provided in the third prompt of Appendix~\ref{prompt}.

\noindent {\bf Unit Test-based Evaluation.}
For Task-3, we assess whether the migrated code preserves the original functionality by running a set of test cases on both the original and migrated implementations.

Specifically, three test cases are executed for each code pair. The outputs of the original code serve as the reference, and the migrated code is considered correct only if it produces identical outputs for all test cases. The accuracy for Task-3 is defined as:
\begin{equation}
\mathrm{Acc}_{\text{Task-3}} = \mathbbm{1}\left[\,\bigwedge_{i=1}^{3}(m_i = o_i)\,\right],
\end{equation}
where $m_i$ is the output of the migrated code and $o_i$ is the output of the original code for the $i$-th test case.

\subsection{Construction Process}

Figure~\ref{fig:process} presents the data curation workflow for~\OurDATA{}, which comprises three stages:

\noindent {\bf Step 1: Data Collection.} We begin by gathering a comprehensive set of functions, along with their associated changes, descriptions, and supported version ranges.

To achieve this, we systematically review version release notes from the official documentation of each package, cataloging all modified functions and detailing their changes across different versions. We also extract functional descriptions and usage information for each function to support subsequent code generation. Since official documentation often omits explicit version compatibility information, we empirically determine the supported version ranges by executing the functions across multiple package versions.

In total, our analysis yields 212 function changes for Python and 114 for Java. The online sources referenced for this collection are listed in Appendix~\ref{source}.

\noindent {\bf Step 2: Code Generation.} Next, we generate original code samples based on the collected data. This step leverages the advanced capabilities of GPT-4: by providing it with the function changes, original function definitions, and usage descriptions, we prompt GPT-4 to generate code that correctly utilizes these functions.

The code generation scenario depends on the type of function change (\textit{i.e.}, \textsc{Old2New} or \textsc{New2Old}):

\begin{itemize}
    \item[(i)] For addition-type changes ($\texttt{None} \rightarrow f_{\text{new}}$), we create \textsc{New2Old} samples. GPT-4 is given the newly introduced function $f_{\text{new}}$ and asked to generate code compatible with the newer environment where $f_{\text{new}}$ exists. The migration target is the version prior to the change, where $f_{\text{new}}$ is unavailable.
    \item[(ii)] For deprecation-type changes ($f_{\text{old}} \rightarrow \texttt{None}$), we create \textsc{Old2New} samples. GPT-4 receives the deprecated function $f_{\text{old}}$ and generates code that runs in the older environment where $f_{\text{old}}$ is still present. The migration target is the version after the change, where $f_{\text{old}}$ has been removed.
    \item[(iii)] For replacement-type changes ($f \rightarrow f'$), we generate both \textsc{Old2New} and \textsc{New2Old} samples. GPT-4 is prompted separately with each function to produce the corresponding code samples for both migration directions.
\end{itemize}

Further details on this process are provided in the first and sixth prompts of Appendix~\ref{prompt}.

\noindent {\bf Step 3: Test Case Generation.} Finally, we construct test cases for each generated code sample to ensure that migrated code preserves functional correctness.

For this, GPT-4 is supplied with both the original code and its functional specification and instructed to generate three test cases. These are executed on the original code to obtain ground-truth outputs, which are then used in Task-3 to assess the correctness of migrated code.

Occasionally, generated test cases may exhibit issues such as invalid input ranges, incorrect formats, or runtime errors, which may arise from either the test cases themselves or defects in the original code. To address this, we iteratively provide GPT-4 with error messages from failed executions, allowing it to refine the test cases. This refinement process is repeated for up to three iterations. If all three test cases execute successfully, the data sample is retained; otherwise, both the test cases and the associated code are discarded.

Details of this step are provided in the fourth prompt of Appendix~\ref{prompt}.

\vspace{-1.7ex}
\section{Experimentation}
\vspace{-1.2ex}
In this section, we present a comprehensive analysis of our experimental setup and results.

\vspace{-1.2ex}
\subsection{Experimental Settings}

\begin{table*}[ht]
\centering
\resizebox{\linewidth}{!}{
\begin{tabular}{lp{1.5cm}p{1.5cm}p{1.5cm}p{1.5cm}p{1.5cm}p{1.5cm}p{1.5cm}p{1.5cm}}
\toprule[1.0pt]

\multirow{2}{*}{\textbf{Base Model}} & \multicolumn{4}{c}{\textbf{Task 1 Locating Function}}  & \multicolumn{4}{c}{\textbf{Task 2 Answering Change}} 

\\ \cmidrule{2-9}  
& \multicolumn{1}{l}{\textbf{Python (easy)}} & \multicolumn{1}{l}{\textbf{Python (hard)}} & \multicolumn{1}{l}{\textbf{Java}} & \multicolumn{1}{l}{\textbf{Avg.}}

& \multicolumn{1}{l}{\textbf{Python (easy)}} & \multicolumn{1}{l}{\textbf{Python (hard)}} & \multicolumn{1}{l}{\textbf{Java}} & \multicolumn{1}{l}{\textbf{Avg.}}\\
\hline
\textsc{GPT-Turbo-3.5}     & \textbf{85.10} & \textbf{32.98} & 80.89  & \textbf{66.32} & \underline{26.01} & 13.09 & 63.28 & 34.13 \\
\textsc{GPT-4o-Mini} & 77.21 & 21.99 & \textbf{84.77} & 61.32  & 18.73 & 6.28 & 68.95 & 31.32  \\
\textsc{GPT-4o}     & 70.71 & 25.65 & 81.19  & 59.18 & 22.22 & \underline{13.61} & \textbf{75.22} & \underline{37.02} \\
\textsc{Llama-3.1-8B}   & 70.71 & 21.99 & 67.16  & 53.29 & 16.16 & 2.09 & 53.13 & 23.79 \\
\textsc{Llama-3.1-70B}   & 75.51 & \underline{29.84} & 81.19  & 62.18 & 22.73 & 8.38 & \textbf{75.22} & 35.44  \\     
\textsc{DeepSeek-v3}     & \underline{78.48} & 26.17 & \underline{82.08}  & \underline{62.24} & \textbf{38.99} & \textbf{16.75} & \underline{70.44} & \textbf{42.06}  \\
\bottomrule[1.0pt]
\end{tabular}}
\caption{Experiment results for Task-1 and Task-2. We \textbf{bold} the best result and \underline{underline} the second-best result. The first two tasks only test general LLMs, because code-specialized LLMs focus more on generating code.}
\vspace{-2.7ex}
\label{tab:res_table2}
\end{table*}

\noindent{\bf Large Models.} 
We conduct experiments on nine different LLMs. 
These include six general LLMs, namely: \textsc{GPT-Turbo-3.5} \cite{ye2023comprehensivecapabilityanalysisgpt3}, \textsc{GPT-4o-Mini} \cite{openai2024gpt4technicalreport}, \textsc{GPT-4o} \cite{openai2024gpt4ocard}, \textsc{Llama-3.1-8B-Instruct} \cite{Dubey2024TheL3}, \textsc{Llama-3.1-70B} \cite{Dubey2024TheL3}, and \textsc{DeepSeek-V3} \cite{Shao2024DeepSeekV2AS}; three code-specialized LLMs:  \textsc{Qwen2.5-Coder-7B-Instruct} \cite{Hui2024Qwen25CoderTR}, \textsc{StarCoder2-15B} \cite{Lozhkov2024StarCoder2A}, and \textsc{Code Llama-34B} \cite{Rozire2023CodeLO}.

\noindent{\bf Evaluation Metrics.} We assess model performance on the three tasks using the following metrics: $\text{Acc}_{\text{Task\_1}}$, $\text{Acc}_{\text{Task\_2}}$, and $\text{Acc}_{\text{Task\_3}}$, as detailed in Section~\ref{metrics}.

For Task-3 (code migration), we further report the Pass@$k$ metric~\cite{Hendrycks2021MeasuringCC}, which quantifies the proportion of examples for which the model produces at least one correct migration within $k$ attempts. Formally,

\begin{equation}
\text{Pass@}k = \mathbb{I}\left[\bigcup_{i=1}^{k} \text{Acc}_{\text{Task\_3}}^{(i)} \right]
\end{equation}

where $\text{Acc}_{\text{Task\_3}}^{(i)}$ is an indicator of whether the $i$-th attempt for Task-3 is successful.

\noindent{\bf Experiment Setup.} 
For all LLMs, we set the generation temperature to 0.7 and limit the maximum output sequence length to 2048 tokens. Proprietary models (e.g., the GPT series) are evaluated via their official APIs. For smaller open-source models such as \textsc{Qwen2.5-Coder-7B-Instruct}, we run inference locally using two RTX 4090 GPUs.
For large-scale open-source models, ~\textit{i.e.,} \textsc{Llama-3.1-70B}, we access them through 
APIs provided by third-party websites \footnote{\url{https://cloud.siliconflow.cn/models}}.

\subsection{Main Experiments}
Table~\ref{tab:res_table2} summarizes the experimental results for Task-1 and Task-2. Below, we discuss the key findings:

\noindent{\bf Overall Performance of Task-1.}  
The average locating success rate ($\text{Acc}_{\text{Task\_1}}$) for the six general LLMs across both Python and Java is 59.76\%. Performance is notably strong on the Python (easy) and Java datasets, with average scores of 74.84\% and 79.53\%, respectively. However, all models struggle on the Python (hard) dataset, which contains multiple incompatible functions per example. For instance, \textsc{Qwen2.5-Coder-7B} achieves only a 15.71\% pass rate in this setting. This drop in performance is primarily due to the models' difficulty in identifying all incompatible functions: when several such functions are present, models often detect only a subset or make incorrect identifications.

Overall, \textsc{GPT-Turbo-3.5} achieves the highest overall success rate on Task-1, with an average of 66.32\%. Its strength is particularly apparent on the Python (hard) dataset, where it attains a 32.98\% pass rate—substantially higher than other models. This suggests that \textsc{GPT-Turbo-3.5} is more adept at comprehensively locating all version-incompatible functions in complex scenarios. While other models may overlook or misclassify some incompatible functions when multiple are present, \textsc{GPT-Turbo-3.5} more consistently identifies a greater proportion, leading to its superior performance. Nevertheless, it does not achieve the top results on Java, which may reflect differences in model proficiency across different programming languages.

\begin{table*}[ht]
\centering
\resizebox{\linewidth}{!}{
\begin{tabular}{lp{1.5cm}p{1.5cm}p{1.5cm}p{1.5cm}p{1.5cm}p{1.5cm}p{1.5cm}p{1.5cm}}
\toprule[1.0pt]
\multirow{3}{*}{\textbf{Base Model}} & \multicolumn{4}{c}{\textbf{Task 3 Migration (\textsc{Old2New})}}  & \multicolumn{4}{c}{\textbf{Task 3 Migration (\textsc{New2Old})}} \\ \cmidrule{2-9} & \multicolumn{2}{c}{\textbf{Python (\textbf{easy})}}  & \multicolumn{2}{c}{\textbf{Python (hard)}}  & \multicolumn{2}{c}{\textbf{Python (easy)}}  & \multicolumn{2}{c}{\textbf{Python (hard)}}   \\ \cmidrule{2-9}
& \multicolumn{1}{c}{\textbf{Pass@1}} & \multicolumn{1}{c}{\textbf{Pass@5}} &  \multicolumn{1}{c}{\textbf{Pass@1}} & \multicolumn{1}{c}{\textbf{Pass@5}} & \multicolumn{1}{c}{\textbf{Pass@1}} & \multicolumn{1}{c}{\textbf{Pass@5}} & \multicolumn{1}{c}{\textbf{Pass@1}} & \multicolumn{1}{c}{\textbf{Pass@5}}\\
\hline

\multicolumn{9}{c}{\textsc{General Large Language Model}} \\ \cmidrule{1-9} 

\textsc{GPT-Turbo-3.5}   & 26.03 & 34.93 & 7.32  & 10.98 & 24.80 & 38.40 & 7.34 & 9.17 \\
\textsc{GPT-4o-Mini} & 30.82 & 49.32 & 15.85 & 26.83 & \underline{29.60} & \textbf{44.00} & 11.93 & 16.51 \\

\textsc{Llama-3.1-8B}  & 23.97 & 28.08 & 8.54 & 10.97 & 20.80 & 24.00 & 7.34 & 11.93    \\
\textsc{Llama-3.1-70B}  & 32.88 & 45.89 & 19.51 & \underline{35.37} & 28.80 & 40.80 & \underline{17.43} & 19.27 \\

\textsc{DeepSeek-v3}  & \underline{41.20} & \underline{54.11} & \underline{20.73} & 29.27 & \underline{29.60} & 39.60 & 14.68 & \underline{23.85}    \\
\textsc{GPT-4o} & \textbf{43.84} & \textbf{59.59} & \textbf{26.83} & \textbf{47.56} & \textbf{31.60} & \underline{43.60} & \textbf{22.94} & \textbf{27.52} \\
 \hline
\multicolumn{9}{c}{\textsc{Code-Specialized Large Language Model}} \\ \cmidrule{1-9}

\textsc{Qwen2.5-Coder-7B}   & \underline{32.19} & \underline{46.58} & \underline{14.63} & 24.39 & \underline{29.20} & 38.00 & 8.26 & 12.84   \\

\textsc{StarCoder2-15B}   & \underline{32.19} & \underline{46.58} & 12.54 & \underline{28.73} & 28.80 & \underline{38.40} & \underline{13.50}& \underline{18.35}   \\

\textsc{Code Llama-34B}   & \textbf{35.62} & \textbf{53.42} &  \textbf{21.95} & \textbf{36.49} & \textbf{29.60} & \textbf{40.80} & \textbf{15.76} & \textbf{21.10}   \\

\bottomrule[1.0pt]
\end{tabular}}
\caption{Experiment results for Task-3, we report the results in two cases: \textsc{Old2New} and \textsc{New2Old}.}
\vspace{-2.7ex}
\label{tab:migration}
\end{table*}

\noindent{\bf Overall Performance of Task-2.} The average success rate ($\text{Acc}_{\text{Task\_2}}$) for the six general LLMs across Python and Java is 33.96\%, which is notably lower than their performance on Task-1. This gap highlights a key limitation: while LLMs can often identify version-incompatible functions, they struggle to recall or reason about the specific details of how those functions have changed across versions. In other words, LLMs are less adept at providing precise, contextually accurate descriptions of function modifications, replacements, or deprecations.

Among all models, \textsc{DeepSeek-v3} stands out with the highest average score of 42.06\%. Its advantage is particularly evident on the Python (easy) dataset, where it achieves 38.99\%. A closer look at the scoring criteria for $\text{Acc}_{\text{Task\_2}}$ (see Section~\ref{metrics}) reveals that \textsc{DeepSeek-v3}'s strength lies in its ability to accurately recall the specific version in which a function change occurred—a capability that most other LLMs lack. This suggests that \textsc{DeepSeek-v3} has either been exposed to more up-to-date or detailed training data, or possesses better mechanisms for temporal reasoning about software evolution.

Conversely, \textsc{Llama-3.1-8B} lags behind, with an average score of only 23.79\%. Its primary weakness is in identifying replacement-type changes: it often fails to specify which function an incompatible one should be replaced with in the target environment. This indicates that smaller or less specialized models may lack the depth of codebase knowledge or the reasoning ability required for nuanced migration scenarios.

\noindent{\bf Overall Performance of Task-3.} Table~\ref{tab:migration} presents the results for Task-3 (code migration). In the \textsc{Old2New} scenario, the average Pass@1 success rate for the nine LLMs is 33.56\% on the easy set and drops to 16.20\% on the hard set. Notably, allowing more attempts substantially boosts performance: Pass@5 rises to 45.5\% (easy) and 26.47\% (hard), indicating that LLMs can often generate a correct migration with additional tries, even if their first attempt fails.

In contrast, the \textsc{New2Old} scenario proves much more challenging. Here, the average Pass@1 and Pass@5 rates are only 12.77\% and 17.30\% (hard set), and additional attempts yield little improvement. This asymmetry suggests that LLMs are more familiar with migrating legacy code to newer environments than the reverse, likely reflecting the distribution of code and documentation in their training data.

Among all models, \textsc{GPT-4o} delivers the strongest performance in the \textsc{Old2New} migration task, achieving a Pass@1 rate of 43.84\% and a Pass@5 rate of 59.59\% on the easy set. This demonstrates its superior ability to synthesize and adapt code for modern environments, likely due to its larger context window, more recent training data, and advanced reasoning capabilities. In contrast, \textsc{GPT-Turbo-3.5}, which excels at locating incompatible functions (Task-1), does not translate this strength into effective code migration: its Pass@1 rate on the hard set is only 7.32\%. This discrepancy highlights that the skills required for identifying incompatibilities and for generating correct, environment-adapted code are distinct. While \textsc{GPT-Turbo-3.5} can assist users in pinpointing problematic functions, it is less reliable for fully automated migration, especially in complex scenarios.





\noindent{\bf Preference of New Functions.} We find that LLMs are more familiar with the new functions compared to the old ones. Our experimental results show that LLMs perform better in the \textsc{Old2New} task compared to the \textsc{New2Old} task. For example, \textsc{GPT-4o} achieves a pass@1 rate of 44.52\% in the \textsc{Old2New} task at easy difficulty, while for \textsc{New2Old} at the same difficulty, it only reaches 28.00\%. A possible reason for this is that the demand for writing code for new environments is more widespread, and during the training process, the proportion of new functions in the training data is higher than that of old functions, leading to this function preference. Furthermore, this trend varies in magnitude across different models. For instance, \textsc{GPT-4o-Mini} shows a smaller performance gap between \textsc{New2Old} and \textsc{Old2New}.

\begin{figure*}[t!]
    \centering
    \includegraphics[width=1\linewidth]{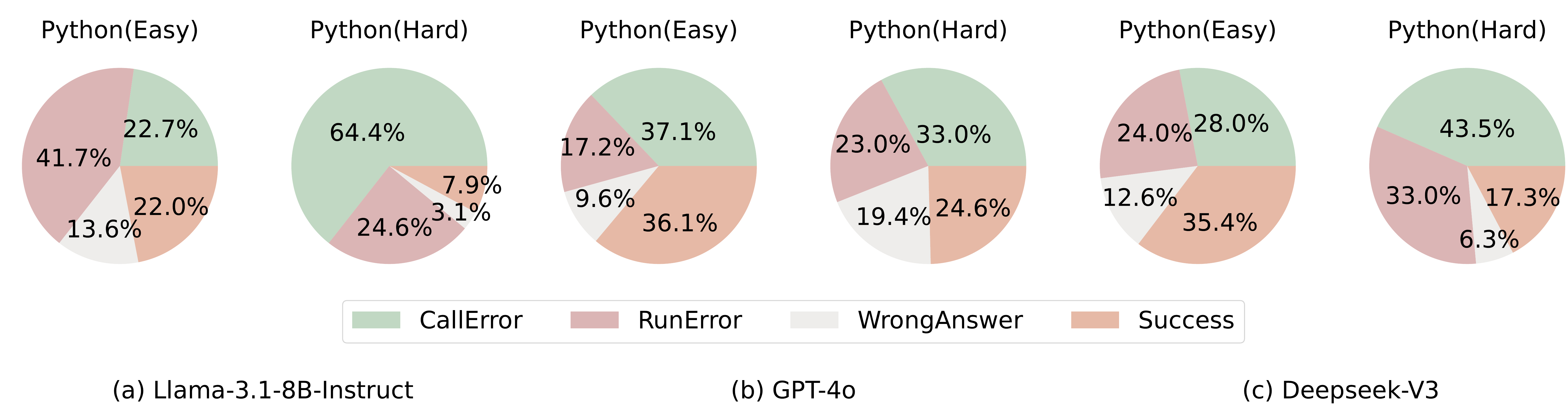}
    \caption{Error Analysis of Code Migration. \texttt{CallError} represents a function 
    where an incompatible the environment is still called. \texttt{RunError} represents 
    that the migrated code enters an infinite loop during execution.  \texttt{WrongAnswer} 
    represents this code runs normally and gets the result, but it is different from the 
    standard answer. We combine the experimental results of \textsc{New2Old} and 
    \textsc{Old2New} in this pie chart.}
    \label{fig:ErrorAnalysis}
    \vspace{-2.7ex}
\end{figure*}


\noindent{\bf Error Analysis for Code Migration.} To better understand the types of errors that occur during code migration, we conduct an error analysis of the failed cases, as illustrated in Figure~\ref{fig:ErrorAnalysis}. We categorize the failures into several types. The most prevalent is \texttt{CallError}, which arises when the generated code still invokes a function that is incompatible with the target environment. For example, 50.8\% of the code produced by \textsc{Llama-3.1-8B} for the Python (\textbf{hard}) migration task fails due to this error. Such failures can occur either because the model does not successfully identify all incompatible functions, or because, even after correctly locating them, it still generates code that calls an incompatible function. Another common error type is \texttt{RunError}, where the code compiles and runs but enters an infinite loop or otherwise fails to terminate in a reasonable time. For instance, 33.0\% of the code generated by \textsc{Deepseek-Chat} failed due to this issue.

Additionally, some migrated code, while calling functions compatible with the environment 
and passing compilation successfully, produces results that deviate from the expected output, 
leading to a \texttt{WrongAnswer}. For instance, 19.4\% of the code generated by \textsc{GPT-4o} 
failed due \texttt{WrongAnswer}.

\begin{figure*}[t!]
    \centering
    \includegraphics[width=0.8\linewidth]{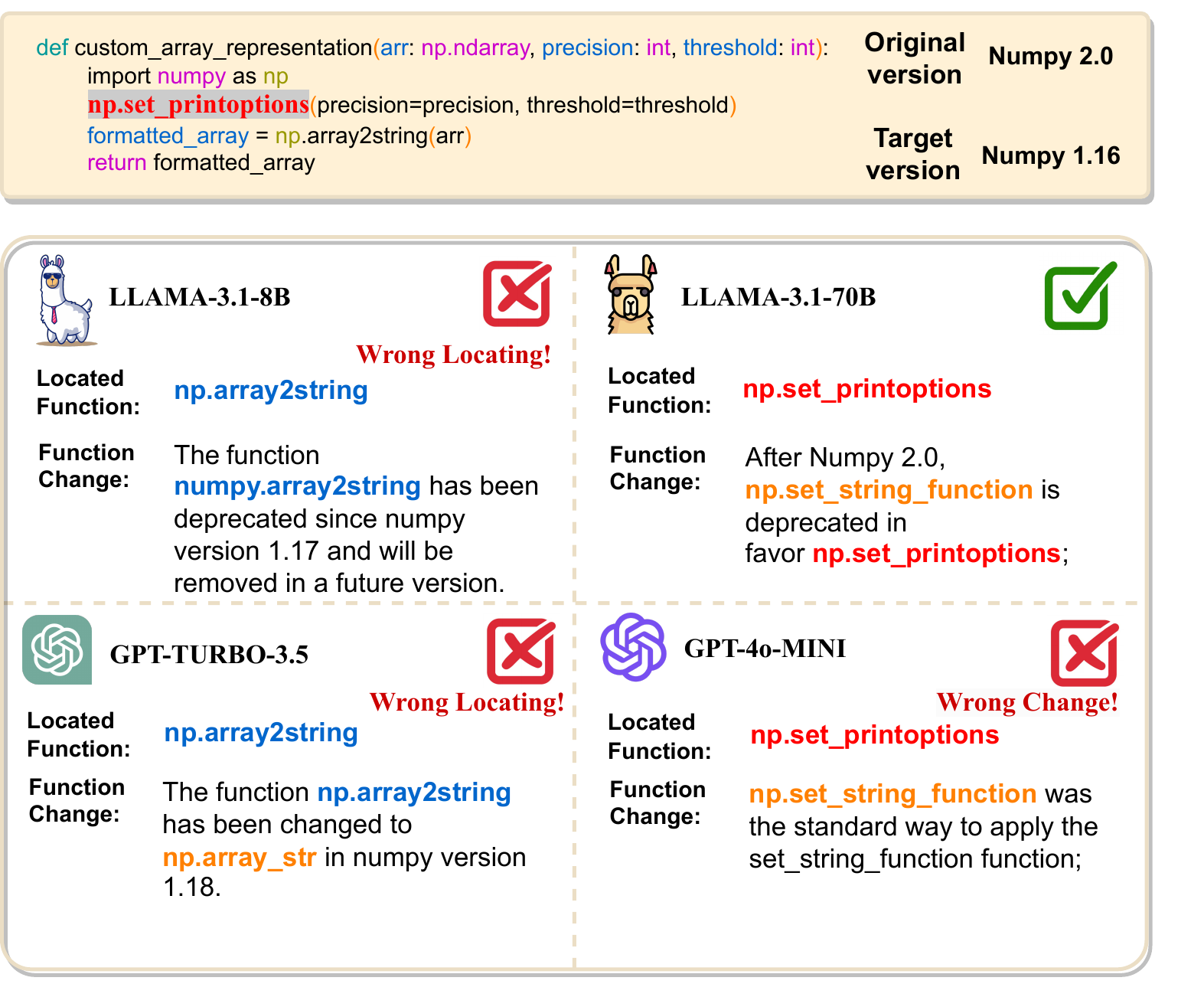}
    \caption{Case Study. We plot an example of \textsc{New2Old} from Python (Easy) datasets and present the response of Task-1 and Task-2 for four LLMs. In this case study, we observe the phenomenon of logical inconsistency, where \textsc{Llama-3.1-8B} and \textsc{GPT-Turbo-3.5} provide function changes that are unrelated to the migration process.}
    \vspace{-2.7ex}
    \label{fig:casestudy}
\end{figure*}

\noindent{\bf Case Studies.} 
The goal of this case study (Figure~\ref{fig:casestudy}) is to analyze how different LLMs perform on the tasks of locating version-incompatible functions and accurately describing their changes, with a focus on their reasoning about version constraints.

Our findings reveal two main types of errors. First, both \textsc{Llama-3.1-8B} and \textsc{GPT-Turbo-3.5} fail to correctly identify the relevant incompatible function. Instead, they focus on \texttt{np.array2string}, providing information about changes introduced in NumPy versions 1.17 and 1.18, even though the target environment is 1.16. Since these changes do not impact functionality in version 1.16, the models' responses are irrelevant for the migration task. This suggests a common failure mode: incorrect reasoning about version ordering, where models conflate changes from later versions with the requirements of an earlier target environment.

In contrast, \textsc{Llama-3.1-70B} and \textsc{GPT-4o-Mini} correctly identify \texttt{np.set\_printoptions} as incompatible with NumPy 1.16. However, \textsc{GPT-4o-Mini} struggles to specify the precise version in which the function change occurred, providing inaccurate version information. This issue—misreporting the version associated with a function change—was frequently observed across our evaluation, highlighting a broader challenge for LLMs in tracking the evolution of library APIs with precision.

Overall, this case study demonstrates that even when models can identify relevant functions, they often fail in reasoning about version boundaries and providing accurate change details, which are critical for reliable code migration.


\vspace{-1.7ex}
\section{Conclusion}
\vspace{-1.7ex}
In this work, we introduced \OurDATA{}, a comprehensive benchmark designed to assess the code migration capabilities of LLMs across different environments. \OurDATA{} encompasses three core tasks: detecting version-incompatible functions, identifying specific function changes, and migrating code to ensure compatibility.

Our evaluation of nine LLMs demonstrates that models are generally more proficient with newer function versions, which poses difficulties for migrating code from newer to older environments (\textsc{New2Old}). Additionally, our error analysis highlights logical inconsistencies, where the changes proposed by models do not always facilitate successful migration. We hope that \OurDATA{} and the insights from our experiments will inspire further research into improving LLM-driven code migration.

\vspace{-1.3ex}
\section*{Limitations}
\vspace{-1.3ex}
\OurDATA{} is relatively small, particularly the Java dataset. Additionally, the language features of Java make it challenging to establish rigorous unit tests. \OurDATA{} currently involves only two programming languages, Python and Java. We plan to add more programming languages in the future.

\vspace{-1.7ex}
\section*{Ethics Statement}
\vspace{-1.3ex}
Throughout our work, we have strictly adhered to ethical standards. The creation of our dataset also complies with open-source regulations, and the data has undergone manual checks to prevent harmful content.

\vspace{-1.7ex}
\section*{Acknowledgements}
\vspace{-1.7ex}
This work is supported in part by the funding BAS/1/1689-01-01, URF/1/4663-01-01, REI/1/5232-01-01, REI/1/5332-01-01, and URF/1/5508-01-01 from KAUST, and funding from KAUST - Center of Excellence for Generative AI, under award number 5940.

\bibliography{anthology,custom}
\bibliographystyle{acl_natbib}

\appendix
\clearpage

\section{Additional Related Work}
\label{relatedApp}
\subsection{Knowledge Editing}
Knowledge editing is an effective way to add the latest knowledge of function changes to LLMs.

The research on knowledge editing for LLMs aims to efficiently modify large model's parameters in order 
to update its knowledge. Most studies in this field focus on editing natural language knowledge. ROME \citep{meng2022locating} and MEMIT \citep{meng2022mass} adopt a locate-then-edit paradigm, where the parameter position of the knowledge is first located, and then the parameter is updated to modify the model's knowledge. Some work adopts a plan-and-solve paradigm~\cite{zhong2023mquake,Cheng2024MultihopQA}, where complex problems are decomposed into the knowledge required for each step, which are then solved one by one. \cite{Zhang2024LocatetheneditFM} proposes a locate-then-edit paradigm to support efficient knowledge editing for multi-hop questions.

There are only a few research attempts on changes 
to function: CodeUpdateArena \cite{Liu2024CodeUpdateArenaBK} introduces a benchmark for updating LLMs 
with new API function knowledge to solve program synthesis tasks. 
CLMEEval \cite{Li2024ModelEF} propose a benchmark for evaluating model editing 
techniques on LLMs4Code, and proposes A-GRACE, an enhanced method for better 
generalization in code knowledge correction. Some of the recent works \cite{Zhou2022DocPromptingGC,Su2024EvoRER,Hsieh2023ToolDE} use retrieval-augmented 
approaches \cite{Lewis2020RetrievalAugmentedGF,Guu2020REALMRL} to provide models 
with code change knowledge for improving code generation.

Note, unlike existing work,~\OurDATA{} does not supply the model with 
contextual knowledge of function changes during evaluation. Instead, we prioritize 
assessing how effectively the model leverages its inherent knowledge of function 
changes to perform code migration.

\section{Prompts for \OurDATA{}}
\label{prompt}
See \hyperref[prompt1]{Prompt1} for the generation of original code of Python language (step-2 of datasets construction).

See \hyperref[prompt2]{Prompt2} for prompt we used in experiment to execute the three tasks of \OurDATA{} (Python).

See \hyperref[prompt3]{Prompt3} for the agent-based evaluation for Python language.

See \hyperref[prompt4]{Prompt4} for the generation of test cases.

See \hyperref[prompt5]{Prompt5} for the improving of test cases.

See \hyperref[prompt6]{Prompt6} for the generation of original code of Java language (step-2 of datasets construction).

See \hyperref[prompt7]{Prompt7} for prompt we used in experiment to execute the three tasks of \OurDATA{} (Java).

See \hyperref[prompt8]{Prompt8} for the agent-based evaluation for Java language.

\clearpage
\onecolumn
\noindent \phantomsection \label{prompt1}
\begin{center}
	\footnotesize
	\begin{tcolorbox}[width=\textwidth,title={\textbf{Prompt 1. Original Code Generation: the Second Step for Dataset Construction of Python language}}]
    
		==== SYSTEM ====\\
		You are a very experienced programmer who is familiar with the usage of many functions and is good at applying them. At the same time, you are thoughtful and creative and like to apply some functions to solve algorithmic problems.\\
        
        First of all, I will give you an existing library function, you will get the function with signatures and functionality, as well as import methods. I hope you can think about the application of this library function according to the description of this library function, be bold and creative, and then write a piece of code that calls this library function, we call this code as solution. 
        This solution is a function, and should be able to solve medium and difficult algorithmic problems, which require "multiple inferences", at least three or four steps to solve, rather than simply calling your library function. There should be no comments in this solution.\\

        Then, design a problem for the solution you generated, with the requirement that others should be able to derive a solution from this problem. Your problem description should focus on the solution's functionality, as well as its inputs and outputs, rather than guiding the step-by-step generation of the solution within the description\\
        You must explicitly specify the data type and dimensionality of each input parameter, as well as the data type and parameters of the output. Your problem description should follow this template: "Please use Python code to implement a function..." Indicate which library is being utilized in the description, but refrain from specifying the exact library function being called. Avoid disclosing any implementation details.\\

        Note: Do not alias when importing; Here's a return template,only output the JSON in raw text. Don't return anything else.\\
        
        \{\\
            solution\_function: The function that you generate. Make sure the code you return is runnable.\\
            solution\_signature: The signature of the function you generated, indicating the input and output. And the name of the solution should derive from its functionality,\\
            problem: Generate a literal description of this function. Describe the data type and dimensions of each input parameter and the data type and dimension of the output.\\
        \}\\

        ==== USER ====\\
        The package name for the new library function is:\\
        \textbf{<PACKAGE>}\\
        The import method is as follows:\\
        \textbf{<IMPORT>}\\
        The signature of the new library function is:\\
        \textbf{<SIGNATURE>}\\
        The feature description of the new library function is:\\
        \texttt{[DOC]}\\
        \textbf{<DOC\_STRING>}\\
        \texttt{[/DOC]}\\
		Note: Do not alias when importing; Only output the JSON in raw text.Don't return anything else. 
	\end{tcolorbox}
    
\end{center}

\noindent \phantomsection \label{prompt2}
\begin{center}
	\footnotesize
	\begin{tcolorbox}[width=\textwidth,title={\textbf{Prompt 2. The Prompt used by LLMs to Execute the Three Tasks of \OurDATA{}   (Python)}}]
    
		==== SYSTEM ====\\
		You're a good assistant, and now you need to help me change the code.\\
		I'll give you a piece of Python code that contains functions that are incompatible with the target environment. You need to locate the incompatible function in this code. Then give the information about the change of this located function: (i) It must include the change type {deprecation/addition/replacement}; (ii) the replaced function(if the change type is replacement); (iii) and the version of this function that changed. Finally return your corrected code, and you only need to fix the incompatible function in the code.\\
		Note that Only output the JSON in raw text. Don't return anything else. And here's an example of what you returned.\\
		\{\\
		\quad ai\_api\_wrong: There is the wrong function in the code because of the version,\\
            \quad ai\_api\_change: 1.The specified function(error function) has changed due to version changes, such as being added in version..., being abandoned in version..., or the calling method has changed;
            2.The replace method is... 3.The version that the function changed is...\\
		\quad code\_fixed: Entire code modified\\
		\}\\

		Here's an example of an answer.\\
		\{\\
		\quad ai\_api\_wrong:             
            numpy.compare\_chararrays.\\
            \quad ai\_api\_change: 1.replacement 2.use numpy.char.compare\_chararrays instead\\ 3.The function numpy.compare\_chararrays has been removed in numpy version 2.0.\\
		\quad code\_fixed: def string\_array\_similarities(strings1, strings2):\\
		\quad\quad result = []\\
		\quad\quad for s1 in strings1:\\
		\quad\quad\quad temp\_result = 0\\
		\quad\quad\quad for s2 in strings2:\\
		\quad\quad\quad\quad length\_diff = abs(len(s1) - len(s2))\\
		\quad\quad\quad\quad comparison = numpy.char.compare\_chararrays(numpy.array(list(s1)), numpy.array(list(s2)), cmp='==', assume\_equal=False)\\
		\quad\quad\quad\quad similarity = numpy.sum(comparison) - length\_diff\\
		\quad\quad\quad\quad temp\_result = max(temp\_result, similarity)\\
		\quad\quad result.append(temp\_result)\\
		\quad\quad return result\\
		\}\\

		==== USER ====\\
		Here's the code you need to identify errors.\\
		\texttt{[CODE]}\\
		\textbf{<CODE>}\\
		\texttt{[/CODE]}\\
		Here's the Python library you need to modify your code.\\
		\texttt{[PACKAGE]}\\
		\textbf{<PACKAGE>}\\
		\texttt{[/PACKAGE]}\\
		Here's the version of above package.\\
		\texttt{[VERSION]}\\
		\textbf{<VERSION>}\\
		\texttt{[/VERSION]}\\
	\end{tcolorbox}
    
\end{center}

\noindent \phantomsection
\begin{center}
	\footnotesize
	\begin{tcolorbox}[width=\textwidth,title={\textbf{Prompt 3. Agent-based Evluation: Task-1 and Task-2 of Python Language} }]
		==== SYSTEM ====\\
		You are a good helper for a human being. I ask another LLM to locate the function in a piece of code that is incompatible with the environment, and its response include the following contents: (i) The located incompatible function ; (ii) information of the changes of the function; (iii) The migrated code after fixing the error. \\
        ...\\

        Please compare the wrong functions returned by the AI and the correct function I give you. If ai\_api\_wrong contains api\_wrong, the judge\_locate\_answer is 1,unless return 0.\\ 
        Compare whether the change of the function returned by the AI and the real change I give you. You can loosely compare the two changes. If they are related or only have a little difference, the judge\_update\_answer is 1. If two changes are absolutely are completely irrelevant, return 0. Remember if judge\_locate\_answer is 0, judge\_update\_answer must be 0.

        Note that Only output the JSON in raw text.    Don't return anything else. And here's an example of what you returned.\\
		\{\\
		\quad judge\_reason: The reason why the AI determines whether it is correct or wrong,\\
		\quad judge\_locate\_answer: \{0/1\}\\
            \quad judge\_update\_answer: \{0/1\}\\
		\}\\

		==== USER ====\\
		Here's the code that lets the AI judge that     there is an error.\\
		\texttt{[CODE]}\\
		\textbf{<CODE>}\\
		\texttt{[/CODE]}\\
		Here are the apis given by LLM that are not suitable for the target environment.\\
		\texttt{[API\_LOCATE\_BY\_LLM]}\\
		\textbf{<API\_LOCATE\_BY\_LLM>}\\
		\texttt{[API\_LOCATE\_BY\_LLM]}\\
            Here’s the information regarding the changes in this API, which was returned by LLM. \\
		\texttt{[CHANGE\_INFORMATION\_BY\_LLM]}\\
		\textbf{<CHANGE\_INFORMATION\_BY\_LLM>}\\
		\texttt{[CHANGE\_INFORMATION\_BY\_LLM]}\\
		Here are the answers.\\
		\texttt{[API\_REFERENCE\_ANSWER]}\\
		\textbf{<API\_REFERENCE\_ANSWER>}\\
		\texttt{[API\_REFERENCE\_ANSWER]}\\
            \texttt{[CHANGE\_INFORMATION\_REFERENCE\_ANSWER]}\\
		\textbf{<CHANGE\_INFORMATION\_REFERENCE\_ANSWER>}\\
		\texttt{[CHANGE\_INFORMATION\_REFERENCE\_ANSWER]}\\
		The version is too high or too low.\\
		\texttt{[VERSION\_ERROR]}\\
		\textbf{<VERSION\_ERROR>}\\
		\texttt{[/VERSION\_ERROR]}\\
	\end{tcolorbox}
    \label{prompt3}
\end{center}

\noindent \phantomsection \label{prompt4}
\begin{center}
	\footnotesize
	\begin{tcolorbox}[width=\textwidth,title={\textbf{Prompt 4. Test Cases Generation (Step-3 of Datasets Construction)} }]
		==== SYSTEM ====\\
		\texttt{\# Role}\\
            A very experienced programmer who is good at algorithmic reasoning and can write high-quality code.\\

        \texttt{\# Responsibilities}\\
Write 3 sets of *high-quality* and *comprehensive* input test data based on the problem description and benchmark code.\\

The specific description of these requirements is as follows:\\

\texttt{\# Problem:}\\
That is, the problem scenario. The type of input data and the range limit of the input data are often given in the problem.\\
(Problem is between "[PROBLEM]" and "[/PROBLEM]")\\

\texttt{\# Benchmark code:}\\
That is, the given callable code, and its parameters are each set of input data to be passed in (Benchmark code is between "[CODE]" and "[/CODE]")\\

\texttt{\# Implementation steps}\\
Please answer the questions strictly according to the above requirements and the following steps:\\

1. Determine the input data\\
- First analyze the problem and the given code to determine the type of input data,\\

2. Final input data group generation\\
Based on step 1, return the string of the input data group\\
- Return format: case1:{}\\

====== Task start =====\\
Below is the given problem and function.\\
        
        ==== USER ====\\
        \texttt{[PROBLEM]}\\
        \textbf{<PROBLEM>}\\
        \texttt{[/PROBLEM]}\\
        \texttt{[CODE]}\\
        \textbf{<CODE>}\\
        \texttt{[/CODE]}\\
	\end{tcolorbox}
\end{center}

\noindent \phantomsection \label{prompt5}
\begin{center}
	\footnotesize
	\begin{tcolorbox}[width=\textwidth,title={\textbf{Prompt 5: Improve test cases quality} }]
		==== SYSTEM ====\\
       \texttt{\# Role}\\
An experienced data tester who is good at writing more accurate and higher quality test case based on error information.\\

\texttt{\# Responsibilities}\\
Adjust the test case group according to the provided executable script and running information, and return the adjusted test cases.\\

\texttt{\# Executable script:}\\
That is, a script that can be compiled and run, and the script code already contains an array of test cases.(BETWEEN "[TARGET\_IMPLEMENTATION]" and "[/TARGET\_IMPLEMENTATION]")\\

\texttt{\# Running information:}\\
That is, the running information of each set of test cases when the function is running, mainly focusing on error information.(BETWEEN "[MESSAGE]" and "[/MESSAGE]")\\

\texttt{[MESSAGE]}\\
"""\\
5.0\\
error:{{function\_node \_\_wrapped\_\_Mul\_device\_/job:localhost/replica:0/task:0/device:CPU:0}} Incompatible shapes: [3,2] vs. [3] [Op:Mul]\\
10.0\\
"""\\
\texttt{[/MESSAGE]}\\

- output:\\
case1:[[1.0, 2.0], [3.0, 4.0]], [0.5, 0.5],\\
case2:[[ -1.0, -2.0], [-3.0, -4.0]], [0.5, 0.5],\\
case3:[[10.0]], [1.0]\\

\texttt{\# Notes}\\
Here, you only need to pay attention to the test cases with running errors. For arrays without error information records, there is no need to adjust.\\

\texttt{\# Implementation steps}\\
Please strictly follow the above requirements and the following steps to answer the questions:\\
1. Test cases extraction and identification\\
-Extract the parameters passed by the calling function from the executable script as the test cases group\\

	\end{tcolorbox}
\end{center}

\begin{center}
	\footnotesize
	\begin{tcolorbox}[width=\textwidth,title={\textbf{Prompt 5: Improve test cases quality} }]
2. Match the test cases group with the corresponding operation information\\
-Pair the test cases input groups in sequence according to the operation results\\
3. Save the test cases group that runs correctly and replace the test cases group that runs incorrectly\\
-Keep the test cases group that runs correctly unchanged\\
-For the test cases group that runs incorrectly, analyze the cause according to the error information, avoid similar errors, and replace them with new test case groups.\\
4. Finally, just return the modified test cases, do not return unnecessary explanations!\\

====== Task start =====\\
Below is the given executable script and running information.\\

        ==== USER ====\\
        \texttt{[TARGET\_IMPLEMENTATION]}\\
        \textbf{<TARGET\_IMPLEMENTATION>}\\
        \texttt{[/TARGET\_IMPLEMENTATION]}\\
        \texttt{[MESSAGE]}\\
        \textbf{<MESSAGE>}\\
        \texttt{[/MESSAGE]}\\
	\end{tcolorbox}
\end{center}

\noindent \phantomsection \label{prompt6}
\begin{center}
	\footnotesize
	\begin{tcolorbox}[width=\textwidth,title={\textbf{Prompt 6: Original Code Generation: the Second Step for Dataset Construction of Java language} }]
		==== SYSTEM ====\\
        You are a very experienced JAVA programmer who is familiar with various library functions of java and is good at applying them. At the same time, you are thoughtful and creative, and like to apply some functions to solve algorithmic problems.\\
    
        First of all, I will specify that you use an old function to complete a class, this function may have been removed in the new JDK. Assuming that I am running in an old JDK environment, please call the function anyway.\\
    
        Then, generate a usage description for your generated code, and I can ask others to be able to generate the code from the problem.\\

        Note: Do not alias when importing; Here's a return template,only output the JSON in raw text. Don't return anything else.\\

        \{\\  
        \quad java\_code: The function that you generate. Make sure the code you return is runnable.\\
        \quad class\_name: The name of the class you generate.\\
        \quad function\_description: The usage description of your generated code.\\
        \}\\

		==== USER ====\\
		The signature of the new library         function is:
            \textbf{<SIGNATURE>}\\

        Note: Do not alias when importing; Only output the JSON in raw text.Don't return anything else.
            
	\end{tcolorbox}
\end{center}

\noindent \phantomsection \label{prompt7}
\begin{center}
	\footnotesize
	\begin{tcolorbox}[width=\textwidth,title={\textbf{Prompt 7: The Prompt used by LLMs to Execute the Three Tasks of \OurDATA{} (Java)} }]
		==== SYSTEM ====\\
		You're a good assistant, and now you need to help me find the error of the code.
        I'll give you a piece of java code that has errors due to a java JDK version mismatch. You need to locate the wrong functions in this code, and explain what version changes have taken place in the function that caused the error you pointed out.\\
        *Note that your answers must be concise, and you only need to point out the mistake directly.*
        
        Here's an example of an answer:\\
        Output: \\
        ai\_api\_wrong: com.sun.javadoc.AnnotatedType\\ 
        ai\_api\_change: The declarations in this package have been superseded by those in the package jdk.javadoc.doclet. For more information, see the Migration Guide in the documentation for that package.

		==== USER ====\\
		Here's the code you need to identify      
            errors.\\
		\texttt{[CODE]}\\
		\textbf{<CODE>}\\
		\texttt{[/CODE]}\\
		Here's the version of the JDK\\
		\texttt{[VERSION]}\\
		\textbf{<VERSION>}\\
		\texttt{[VERSION]}\\
            
	\end{tcolorbox}
\end{center}

\noindent \phantomsection \label{prompt8}
\begin{center}
	\footnotesize
	\begin{tcolorbox}[width=\textwidth,title={\textbf{Prompt 8: Agent-based Evluation: Task-1 and Task-2 of Java Language} }]
		==== SYSTEM ====\\
		You are a good helper for a human being. I ask another LLM to locate the function in a piece of code that is incorrectly called because of the JDK version mismatch, and its response include the following contents: (i) The located incompatible function ; (ii) information of the changes of the function; (iii) The migrated code after fixing the error.\\

        Please compare the wrong functions returned by the AI and the correct functions I give you. If api\_locate\_by\_llm contains api\_reference\_answer, the judge\_locate\_answer is 1,unless return 0. \\
        Compare whether the change of the function returned by the AI and the real change I give you. You can loosely compare the two changes. If they are related or only have a little difference, the judge\_update\_answer is 1. If two changes are absolutely are completely irrelevant, return 0. Remember if judge\_locate\_answer is 0, judge\_update\_answer must be 0.

        \{\\
	\quad judge\_reason: The reason why the AI            determines whether it is correct or wrong,\\
	\quad judge\_locate\_answer: \{0/1\}\\
        \quad judge\_update\_answer: \{0/1\}\\
	\}\\

		==== USER ====\\
		Here's the code that lets the AI judge that     there is an error.\\
		\texttt{[CODE]}\\
		\textbf{<CODE>}\\
		\texttt{[/CODE]}\\
		Here's the  wrong apis that the AI returned.\\
		\texttt{[API\_LOCATE\_BY\_LLM]}\\
		\textbf{<API\_LOCATE\_BY\_LLM>}\\
		\texttt{[API\_LOCATE\_BY\_LLM]}\\
            Here's the change of the wrong apis that    the AI returned.\\
		\texttt{[CHANGE\_INFORMATION\_BY\_LLM]}\\
		\textbf{<CHANGE\_INFORMATION\_BY\_LLM>}\\
		\texttt{[CHANGE\_INFORMATION\_BY\_LLM]}\\
		Here are the answers.\\
		\texttt{[API\_REFERENCE\_ANSWER]}\\
		\textbf{<API\_REFERENCE\_ANSWER>}\\
		\texttt{[API\_REFERENCE\_ANSWER]}\\
            \texttt{[CHANGE\_INFORMATION\_REFERENCE\_ANSWER]}\\
		\textbf{<CHANGE\_INFORMATION\_REFERENCE\_ANSWER>}\\
		\texttt{[CHANGE\_INFORMATION\_REFERENCE\_ANSWER]}\\
		The version is too high or too low.\\
		\texttt{[VERSION\_ERROR]}\\
		\textbf{<VERSION\_ERROR>}\\
		\texttt{[/VERSION\_ERROR]}\\
            
	\end{tcolorbox}
\end{center}

\clearpage

\section{\OurDATA{} (Additional Details)}
 
\subsection{Datsets Statistics}
\label{statistics}

\begin{table}[ht]
\centering
\resizebox{\linewidth}{!}{
\begin{tabular}{ccccccccc}
\toprule[1.0pt]
 \textbf{Datasets} & \textbf{jdk.nashorn} & \textbf{org.xml} & \textbf{com.sun} &  \textbf{java.applet} &  \textbf{java.beans}  &  \textbf{java.rmi} &  \textbf{java.util}&  \textbf{java.security}\\
\midrule
Java &188 &9 &86 &9 &3 &15 &7 &18\\
\bottomrule[1.0pt]
\end{tabular}}
\caption{Statistics on the number of changes across different Java packages.}
\label{tab:dataset3-single-column}
\end{table}

\begin{table}[ht]
\centering
\resizebox{\linewidth}{!}{
\begin{tabular}{cccccccccccc}
\toprule[1.0pt]
 \textbf{Datasets} & \textbf{numpy} & \textbf{python} & \textbf{math} &  \textbf{re} &  \textbf{os}  &  \textbf{random} &  \textbf{itertools} &  \textbf{torch} &  \textbf{tensorflow} &  \textbf{pandas} &  \textbf{csv}\\
\midrule
\textbf{Python (easy)} &39 &26 &51 &5 &34 &3 &15 &21 &154 &46 &2\\
\textbf{Python (hard)} &20 &- &- &- &- &- &- &21 &115 &35 &-\\
\bottomrule[1.0pt]
\end{tabular}}
\caption{Statistics on the number of changes across different Python packages.}
\label{tab:dataset4-single-column}
\end{table}

\begin{table}[ht]
\centering
\resizebox{0.5\linewidth}{!}{
\begin{tabular}{llll}
\toprule[1.0pt]
\textbf{Package}  & \textbf{Replacement} & \textbf{Deprecation} & \textbf{Addition} \\
\midrule
\multirow{1}{*}{\textbf{numpy}}  &2  &8  &-  \\
\multirow{1}{*}{\textbf{pandas}}  &-  &12  &13 \\  
\multirow{1}{*}{\textbf{tensorflow}}   &87 &2  &2 \\
\multirow{1}{*}{\textbf{python}}   &9 &7  &7 \\
\multirow{1}{*}{\textbf{math}}   &- &1  &17 \\
\multirow{1}{*}{\textbf{re}}   &- &-  &2 \\
\multirow{1}{*}{\textbf{os}}   &- &-  &14 \\
\multirow{1}{*}{\textbf{random}}   &- &-  &2 \\
\multirow{1}{*}{\textbf{csv}}   &- &-  &1 \\
\multirow{1}{*}{\textbf{itertools}}   &- &-  &5 \\
\multirow{1}{*}{\textbf{torch}}   &- &5  &5 \\
\midrule
\multirow{1}{*}{\textbf{total}}   &98 &35  &79 \\
\bottomrule[1.0pt]
\end{tabular}}
\caption{Statistics of the number of three types of function changes across different packages of python language.}
\label{tab:functionchanges}
\end{table}

\subsection{Data Collection Source}
\label{source}
\begin{table}[htbp]
\centering
\label{tab:url_table}
\begin{tabular}{|>{\raggedright\arraybackslash}p{6cm}|p{8cm}|}
\hline
\textbf{URL} & \textbf{Description} \\
\hline
\url{https://github.com/pytorch/pytorch/releases} & Sources for collecting changes related to the PyTorch library.\\
\hline
\url{https://numpy.org/doc/2.0/release/2.0.0-notes.html#changes} &Sources for collecting changes related to the Numpy library. \\
\hline
\url{https://docs.oracle.com/en/java/javase/11/docs/api/deprecated-list.html} & Sources for collecting changes related to the Java library. \\
\hline
\url{https://docs.python.org/zh-cn/3/library/random.html} & Sources for collecting changes related to the random library. \\
\hline
\url{https://github.com/tensorflow/tensorflow/releases/tag/v2.0.0} & Sources for collecting changes related to the tensorflow library. \\
\hline
\url{https://docs.python.org/zh-cn/3/library/itertools.html} & Sources for collecting changes related to the itertools library. \\
\hline
\hline
\end{tabular}
\caption{The URL for collecting data in step 1 of the data construction process.}
\end{table}

\clearpage

\begin{figure}
    \centering
    \includegraphics[width=1\linewidth]{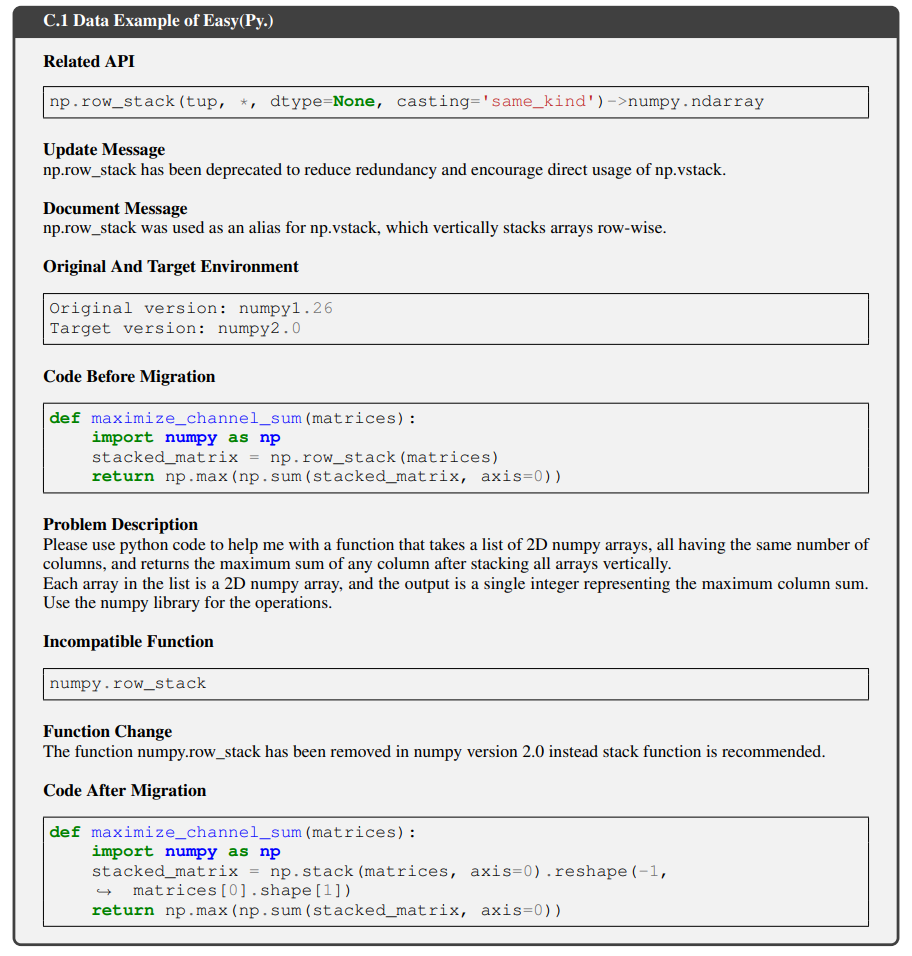}
\end{figure}

\begin{figure}
    \centering
    \includegraphics[width=1\linewidth]{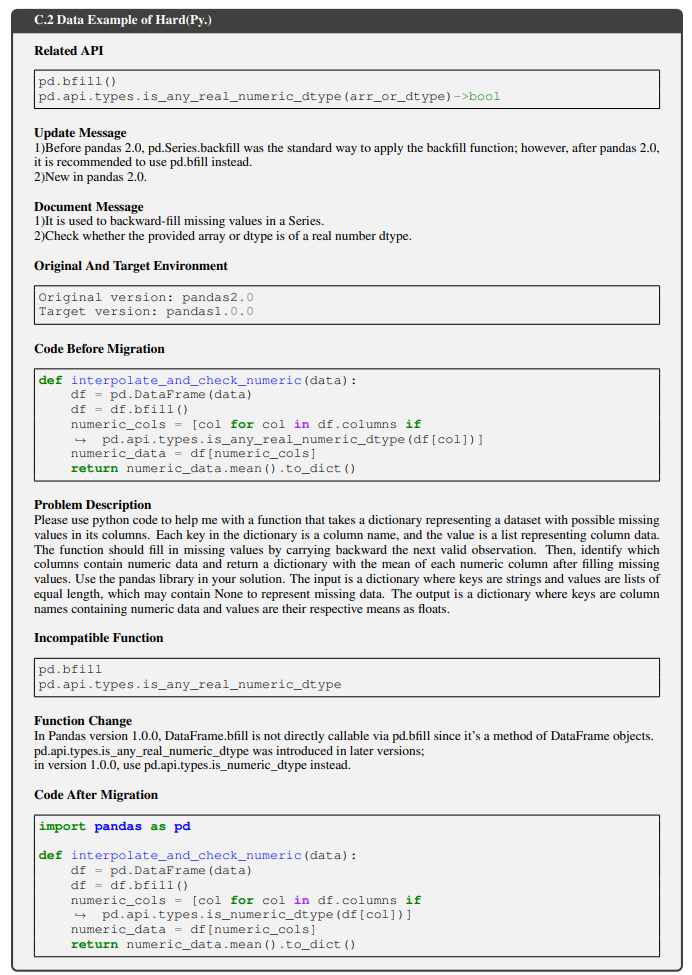}
\end{figure}

\begin{figure}
    \centering
    \includegraphics[width=1\linewidth]{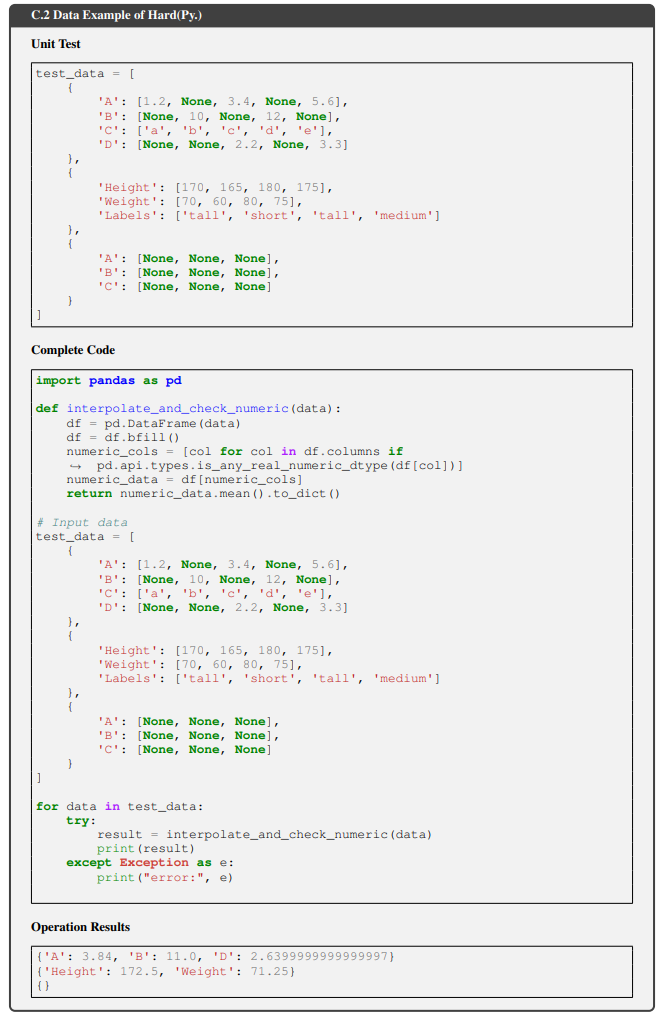}
\end{figure}

\begin{figure}
    \centering
    \includegraphics[width=1\linewidth]{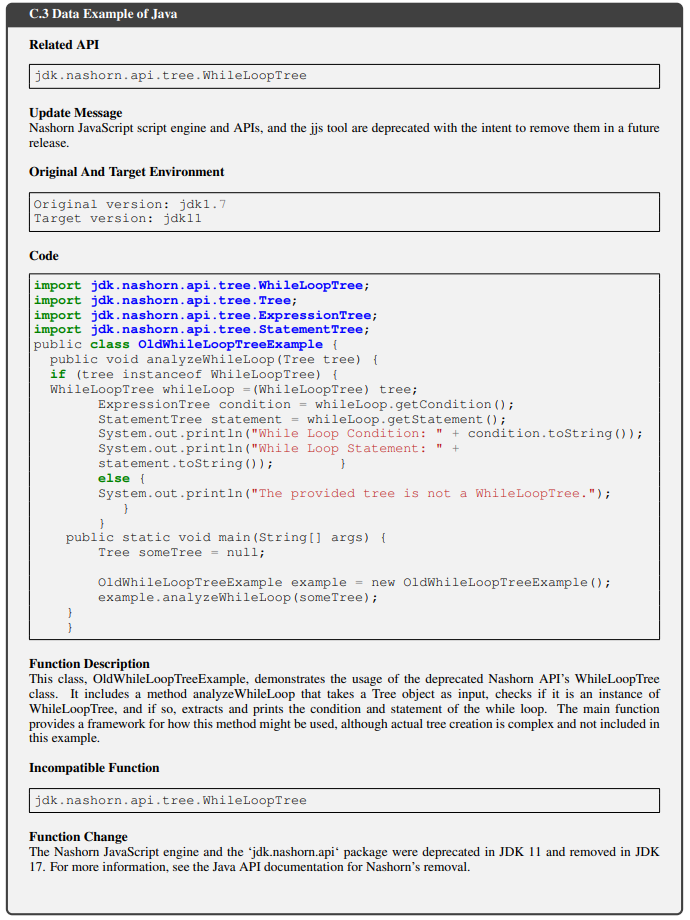}
\end{figure}

\end{document}